\documentclass{elsarticle}
\usepackage{enumerate}
\usepackage{amsmath,slashed,tensor,amssymb,epsfig,amsthm,bm,graphicx,graphics,slashed}
\usepackage[caption=false]{subfig}
\usepackage{hyperref}
\usepackage[utf8]{inputenc}
\usepackage[top=1.5cm, bottom=1.5cm, left=2cm, right=2cm]{geometry}

\usepackage{makecell}
\usepackage{microtype}
\usepackage[font=footnotesize,labelfont=bf]{caption}

\newcommand{\be}{\begin{equation}}\newcommand{\ee}{\end{equation}}
\newcommand{\bea}{\begin{eqnarray}}\newcommand{\eea}{\end{eqnarray}}
\newcommand{\brr}{\begin{array}}\newcommand{\err}{\end{array}}
\newcommand{\bit}{\begin{itemize}}\newcommand{\eit}{\end{itemize}}
\newcommand{\ben}{\begin{enumerate}}\newcommand{\een}{\end{enumerate}}

\newcommand{\bbm}{\begin{bmatrix}}\newcommand{\ebm}{\end{bmatrix}}
\newcommand{\ba}{\begin{array}}
\newcommand{\ea}{\end{array}}
\newcommand{\G}{\textbf}

\newtheorem{mydef}{Definition}
\newtheorem{Lemma}{Lemma}
\newtheorem{theorem}{Theorem}
\newcommand{\bd}{\begin{mydef}} \newcommand{\ed}{\end{mydef}}
\newcommand{\bthe}{\begin{theorem}} \newcommand{\ethe}{\end{theorem}}
\newcommand{\ble}{\begin{Lemma}} \newcommand{\ele}{\end{Lemma}}

\def\ha{\frac{1}{2}}

\def\lf{\left}

\def\non{\nonumber}

\def\ri{\right}
\def\al{\alpha}\def\bt{\beta}
\def\de{\delta}

\def\1{{_{1}}}\def\2{{_{2}}}

\def\noHe0{:\;\!\!\;\!\!:H_e(0):\;\!\!\;\!\!:}
\def\noHm0{:\;\!\!\;\!\!:H_\mu(0):\;\!\!\;\!\!:}

\def\lf{\left}

\def\non{\nonumber}

\def\ri{\right}

\def\al{\alpha}\def\bt{\beta}
\def\de{\delta}

\def\1{{_{1}}}\def\2{{_{2}}}

\begin{document}

\title{Remarks on $SU_q(2)$ fermions}

\author[ipnp]{L.~Smaldone}
\ead{smaldone@ipnp.mff.cuni.cz}

\address[ipnp]{Institute of Particle and Nuclear Physics, Faculty  of  Mathematics  and  Physics, Charles  University, V  Hole\v{s}ovi\v{c}k\'{a}ch  2, 18000  Praha  8,  Czech  Republic.}

\begin{abstract}
We show that most of the applications of $SU_q(2)$ fermions to statistical mechanics and quantum field theory, previously discussed in literature, are based on a wrong statement about the connection between deformed and undeformed fermion operators. Then we exclude various classes of ansatz and we put some constraints about the form of such relation.
\end{abstract}

\vspace{-1mm}

\maketitle
\section{Introduction}
\label{sec:Introduction}
Quantum groups, originally introduced in connection with Quantum Inverse Scattering Method and Yang--Baxter equation \cite{Takhtajan:1989gip}, nowadays represent a widespread mathematical framework: their applications cover various areas of physics, including condensed matter physics \cite{PhysRevLett.121.255302} and quantum gravity \cite{Major_1996, Noui:2002ag, Fairbairn:2010cp, Bianchi:2011uq,Jalalzadeh:2017jdo,Acquaviva:2021aau}.

In the early days of quantum groups, $q$-deformed bosons \cite{Biedenharn:1989jw,Macfarlane:1989dt} and fermions \cite{Ng:1989vk,Chaichian:1989rq,Chaichian:1990ic,Hayashi:1990tf,Frappat_1991} were introduced. In particular, the problem of defining $q$-deformed fermions is cumbersome \cite{Jing:1991rg,Solomon_1994,Wang:1994dh} and many proposals were considered in literature \cite{Viswanathan_1992,Chaichian:1993jk,Arik_2001,Algin_2011}.  One of the main reasons of interest in these subjects, is the possibility of defining new statistics, which generalize the usual Bose--Einstein and Fermi--Dirac distributions \cite{Chaichian:1993jk,Algin:2001mj,ALGIN20082767,Algin_2011}. 

In parallel with such developments, the idea of looking at quantum groups as symmetries was pursued \cite{CHANG1995137}. In particular, one can write down deformations of usual commutation (anticommutation) relations for bosons (fermion), which are invariant under linear transformations belonging to quantum groups as $SU_q(N)$ \cite{Pusz:1989zz,PUSZ1989349,Kempf:1992re,Ubriaco:1993qk}, $GL_{p,q}(2)$ \cite{Jellal2002THERMODYNAMICPO,Algin:2003gl} and $SU_{p,q}(2)$ \cite{Algin:2003gl,Algin:2005pq}. Noticeably, formulas connecting deformed boson and fermion operators with usual (undeformed) ones were derived \cite{UBRIACO1996205,UBRIACO19981}, and applications to statistical mechanics \cite{Jellal2002THERMODYNAMICPO, Algin:2003gl,Algin:2005pq, UBRIACO1996205, Ubriaco:1996xd,PhysRevE.55.291,PhysRevE.57.179,PhysRevE.58.4191} and quantum field theory (QFT) \cite{Tim_teo_1999,Timoteo:1999ga, Timoteo:2006jn} were proposed. Probably, the most important application of $SU_q(N)$ bosons is the derivation of the so-called \emph{generalized uncertainty principle (GUP)} \cite{Kempf:1993bq}, which is a very active area of research \cite{Iorio:2019wtn,Buoninfante:2020cqz,Petruzziello:2020wkd}.

In this paper we show that the above mentioned relations between $SU_q(2)$ fermions and usual fermion operators, only hold for the simplest case of two degrees of freedom (dof) systems. In doing so, we exclude various classes of possible solutions and we propose some ansatz and constraints. 

In Section \ref{suq21} we will review $SU_q(2)$ fermions for two dof. In Section \ref{su2qm} we will analyze the general case of systems with many ($>2$) dof. Finally, in Section \ref{conc} we present our conclusions.
\section{$SU_q(2)$ fermions} \label{suq21}

Let us consider an element of $SU_q(2)$ \cite{Takhtajan:1989gip,Ubriaco:1993qk}
\be
T = \bbm a & b \\ c & d \ebm \, , 
\ee
where the matrix elements satisfy
\bea
a \, b & = & q^{-1} \, b \, a \, , \quad a \, c \ = \ q^{-1} \, c \, a \, ,  \\[2mm]
b \, c & = & c \, b\, ,  \quad d \, c \ = \ q \, c \, d \, , \\[2mm]
d \, b & = & q \, b \, d \, ,  \quad d \, a-a \, d \ = \ (q-q^{-1})\, b \, c \, , \\[2mm]
\mathrm{det}_q T & \equiv & a \, d - q^{-1} \, b \, c \ = \ 1 \, ,  
\eea
with the conjugate matrix
\be
T^\dag \ = \ \bbm d & -q \, b \\ -q^{-1} \, c & a \ebm \, ,
\ee
and $q \in \mathbb{R}$.

Given a doublet of deformed fermion operators $ \G a \ \equiv \bbm a_1 & a_2\ebm^t$, the algebra defined by
\bea \label{acom1}
\lf\{a_1,a_1^\dag\ri\} & = & 1-\lf(1-q^{-2}\ri) \, a^\dag_2 \, a_2 \ = \ q^{-2 \tilde{N}_2} \, , \\[2mm] \label{acom2}
\lf\{a_2,a_2^\dag\ri\} & = & 1 \, ,  \\[2mm] \label{acommix}
a_1 \, a_2 & = & -q \, a_2 \, a_1 \, , \qquad a^{\dag}_1 \, a_2 \ = \ -q \, a_2 \, a_1^\dag \, , \\[2mm] \label{acomt}
\lf\{a_1,a_1\ri\} & = & \lf\{a_2,a_2\ri\} \ = \ 0 \, ,
\eea
is invariant under a linear transformation of the form $\G a' = T \G a$ \cite{Ubriaco:1993qk,UBRIACO1996205}. Here we defined $\tilde{N}_j \equiv a^\dag_j \, a_j $. $a_j,a_j^\dag$ are known as $SU_q(2)$ fermion operators \cite{UBRIACO1996205}.

Remarkably, deformed operators can be written as \cite{UBRIACO1996205}:
\bea \label{ut11}
a_1 & = & A_1 \, (1+\lf(q^{-1}-1\ri) \, N_2) = A_1 \, q^{-N_2}  \, , \\[2mm]
a_1^\dag & = & A_1^\dag \, (1+\lf(q^{-1}-1\ri) \, N_2) = A^\dag_1 \, q^{-N_2} \, , \\[2mm]
a_2 & = & A_2 \, , \qquad a_2^\dag \ = \ A_2^\dag \, , \label{ut21}
\eea
where
\bea \label{sacom1}
\lf\{A_j,A_j^\dag\ri\} & = & 1 \, , \\[2mm]
\lf\{A_i,A_j^\dag\ri\} & = & 0 \, , \qquad i \neq j \, , \\[2mm]
A_J^2 & = & 0 \, , \qquad i,j=1,2 \, ,
\eea
are the usual fermion ladder operators and we introduced the number operators $N_j \equiv A_j^\dag \, A_j$. Moreover, we used that $N_j^2=N_j$.

It is easy to verify Eqs.\eqref{ut11}-\eqref{ut21}. In fact, Eqs.\eqref{acom2},\eqref{acommix} follow trivially. In order to get Eq.\eqref{acom1}, we write
\be
\lf\{a_{1},a^\dag_{1}\ri\}\ = \ q^{-2 N_2} \, \lf\{A_{1},A^\dag_{1}\ri\} \ = \ q^{-2 \tilde{N}_2} \, .
\ee
On the r.h.s. we used that $[A_1,N_2]=[A_1^\dag.N_2]=0$, Eq.\eqref{sacom1} and $N_{2}=\tilde{N}_2$. 

In order to prove Eq.\eqref{acommix} we use that\footnote{To prove this formula, let us put $q=e^s$. Then we consider the function 
$
f(s) \ \equiv \ e^{-s \, N_{2}} \,  A_{2} \, e^{s \, N_{2}} \, .
$
We can easily find that
\be
f'(s) \ \equiv \ - \, e^{-s \, N_{2}} \,  \lf[N_{2} \, , \,  A_{2}\ri] \, e^{s \, N_{2}} \ = \ e^{-s \, N_{2}} \, A_{2} \, e^{s \, N_{2}} \ = \ f(s) \, .
\ee
The solution is then
$
f(s) \ = \ C \, e^s \, = \, C \, q \, .
$
The constant is fixed by the boundary condition $f(0)=A_{2}$. Then
$
q^{-N_{2}} \,  A_{2} \, q^{N_{2}} \ = \ q \, A_{2}  \, .
$
which is equivalent to Eq.\eqref{bch}. }
\be \label{bch}
q^{-N_{2}} \,  A_{2}=q \, A_{2} \, q^{-N_{2}} \, .
\ee
Then
\be
a_1 \, a_2 \ = \ A_1 \, q^{-N_2} \, A_2 \ = \ q \, A_1 \, A_2 \, q^{-N_2} \ = \ -q \, a_2 \, a_1 \, ,  
\ee
and
\be
a^\dag_1 \, a_2 \ = \ A^\dag_1 \, q^{-N_2} \, A_2 \ = \ q \, A^\dag_1 \, A_2 \, q^{-N_2} \ = \ -q \, a_2 \, a^\dag_1 \, . 
\ee
In the next section we will show that such simple arguments cannot be trivially generalized.
\section{$SU_q(2)$ fermions: many degrees of freedom} \label{su2qm}
When we extend such considerations to systems with many dof\footnote{Here we will use a QFT language, with a `'momentum'' index $\G k$.}, the $SU_q(2)$ invariant fermion-algebra is \cite{Ubriaco:1993qk}:
\bea \label{rd}
\lf\{a_{1, \G k},a_{1,\G k'}^\dag\ri\} & = & \de_{\G k \, \G k'}-\lf(1-q^{-2}\ri) \, a^\dag_{2,\G k'} \, a_{2,\G k} \, , \\[2mm] \label{rs2}
\lf\{a_{2,\G k},a_{2,\G k'}^\dag\ri\} & = & \de_{\G k \, \G k'} \, ,  \\[2mm]
a_{1,\G k} \, a_{2,\G k'} & = & -\al \, a_{2,\G k'} \, a_{1,\G k}+\bt \, a_{2,\G k} \, a_{1,\G k'} \, , \label{crel} \\[2mm] \label{crel1}
a_{2,\G k} \, a_{1,\G k'} & = & -\al \, a_{1,\G k'} \, a_{2,\G k}-\bt \, a_{1,\G k} \, a_{2,\G k'} \, , \\[2mm]
a^{\dag}_{1,\G k} \, a_{2,\G k'} & = & -q \, a_{2,\G k'} \, a_{1,\G k}^\dag \, , \label{rs} \\[2mm]
\lf\{a_{1,\G k},a_{1,\G k'}\ri\} & = & \lf\{a_{2,\G k},a_{2,\G k'}\ri\} \ = \ 0 \, , \label{rs1}
\eea
with $\al=\ha(q+q^{-1})$, $\bt=\ha(q^{-1}-q)$. 

The extension of Eqs. \eqref{ut11}-\eqref{ut21} is an highly non-trivial problem. In the literature, where $SU_q(2)$ fermions were studied in statistical mechanics \cite{UBRIACO1996205, Ubriaco:1996xd,PhysRevE.55.291,PhysRevE.57.179,PhysRevE.58.4191} and to QFT \cite{Tim_teo_1999,Timoteo:1999ga, Timoteo:2006jn}, it is assumed that
\bea \label{ut1}
a_{1,\G k} & = & A_{1,\G k} \, (1+\lf(q^{-1}-1\ri) \, N_{2,\G k}) \ = \ A_{1,\G k} \, q^{-N_{2,\G k}}  \, , \\[2mm]
a_{1,\G k}^\dag & = & A_{1,\G k}^\dag \, (1+\lf(q^{-1}-1\ri) \, N_{2,\G k}) \ = \ q^{-N_{2,\G k}} \, A^\dag_{1,\G k}  \, , \\[2mm]
a_{2,\G k} & = & A_{2,\G k} \, , \qquad a_{2,\G k}^\dag \ = \ A_{2,\G k}^\dag \, , \label{ut2}
\eea
where
\bea \label{rds}
\lf\{A_{i, \G k},A_{i,\G k'}^\dag\ri\} & = & \de_{\G k \, \G k'} \, ,  \\[2mm]
\lf\{A_{1, \G k},A_{2,\G k'}^\dag\ri\} & = & \lf\{A_{i, \G k},A_{j,\G k'}\ri\} \ = \ 0 \, , \quad i \neq j =1,2 \, .\label{rs1s}
\eea
and we have defined $N_{j,\G k} \equiv A_{j,\G k}^\dag \, A_{j,\G k}$. We will now prove, by generalizing the arguments of the previous section, that Eqs.\eqref{ut1}-\eqref{ut2} are wrong, because they are incompatible with Eqs.\eqref{rd}-\eqref{rs1}.

 Let us, in fact, consider
\be
a^{\dag}_{1,\G k} \, a_{2,\G k'} \ = \ A^{\dag}_{1,\G k} \, q^{-N_{2,\G k}} \,  A_{2,\G k'} \, .
\ee
In order to proceed we note that, for $\G k= \G k'$, we can use a generalization of Eq.\eqref{bch}
\be \label{bchk}
q^{-N_{2,\G k}} \,  A_{2,\G k}=q \, A_{2,\G k} \, q^{-N_{2,\G k}} \, ,
\ee
which can be proved in the same way as in the previous case. We thus get
\be
a^{\dag}_{1,\G k} \, a_{2,\G k} \ = \ -q A_{2,\G k} \, q^{-N_{2,\G k}} \,  A^\dag_{1,\G k} \ = \ -q \, a_{2,\G k} \, a^\dag_{1,\G k} \, .
\ee
However, for $\G k \neq \G k'$, $[N_{2,\G k},A_{2,\G k'}]=0$. Then
\be
a^{\dag}_{1,\G k} \, a_{2,\G k'} \ = \  A^{\dag}_{1,\G k} \,  A_{2,\G k'} \, q^{-N_{2,\G k}} \ = \ -A_{2,\G k'} \, A^{\dag}_{1,\G k} \, q^{-N_{2,\G k}} \, , 
\ee
namely 
\be
a^{\dag}_{1,\G k} \, a_{2,\G k'} \ = \ -a_{2,\G k'}\, a^{\dag}_{1,\G k} \, , 
\ee
which is not the same as Eq.\eqref{rs} and is not $SU_q(2)$ invariant. Actually, any prescription of the form 
\be
a_{1,\G k} \ = \ A_{1,\G k} \, F(N_{2,\G k};q) \, , \qquad a_{2,\G k} \ = \ A_{2,\G k} \, ,
\ee
where $F$ depends on $N_{2,\G k}$ and not on the total number operator, and so that $F=q^{-2 N_2}$ for two dof, cannot work. In fact 
\be
a^{\dag}_{1,\G k} \, a_{2,\G k'} \ = \  F(N_{2,\G k};q) \, A^{\dag}_{1,\G k} \,  A_{2,\G k'} \ = \ -A_{2,\G k'} \,  \,  \, F(N_{2,\G k};q) \,  A^{\dag}_{1,\G k} \ = \ - a_{2,\G k'}\, a^{\dag}_{1,\G k} \, , 
\ee
for $\G k \neq \G k'$.

A proposal, which is compatible with Eqs.\eqref{rs2},\eqref{rs},\eqref{rs1}, is
\bea \label{ut1s}
a_{1,\G k} & = &  A_{1,\G k} \, q^{-N_{2}}  \, , \\[2mm]
a_{1,\G k}^\dag & = &  q^{-N_{2}} \, A^\dag_{1,\G k}  \, , \\[2mm]
a_{2,\G k} & = & A_{2,\G k}    \, , \\[2mm]
a_{2,\G k}^\dag & = &  A^\dag_{2,\G k}  \, , \label{ut2s}
\eea
with $N_2 \ \equiv \ \sum_\G k \, N_{2,\G k}$. These correctly reduce to Eqs.\eqref{ut11}-\eqref{ut21} in the case of two dof. However, this prescription fails to give back Eq.\eqref{rd}. In fact, it is easy to check that:
\be
\lf\{a_{1,\G k},a^\dag_{1,\G k'}\ri\} \ = \ \de_{\G k \, \G k'} \, q^{-2 N_2} \, .
\ee
Note that this is correct only in the case of two dof (see Eq.\eqref{acom1}). Note that, in general, any prescription of the form
\be
a_{1,\G k} \ = \ A_{1,\G k} \, F(N_{2};q) \, , \qquad a_{2,\G k} \ = \ A_{2,\G k} \, ,
\ee
where $F$ depends only on the total number operator, suffers of the same pathology. In fact
\be
\lf\{a_{1,\G k},a^\dag_{1,\G k'}\ri\} \ = \ \de_{\G k \, \G k'} \, F^2(N_{2};q)  \, ,
\ee
which can fit the right result only in the case of two dof. Note that the same proof also holds for $F(\hat{N}_{2};q) $, with $\hat{N}_2 \equiv \prod_\G p N_{2,\G p}$.

We now try to fix some constraints. A plausible ansatz could be
\be
a_{1,\G k} \ = \ \sum_{\G p} \, A_{1,\G p} \, F^\dag_{\G k \G p}  \, , 
\ee
where $F_{\G k \G p}$ can be either $F\lf(A^\dag_{2,\G p} A_{2,\G k};q\ri)$, $F\lf(A^\dag_{2,\G k} A_{2,\G p};q\ri)$ or $F\lf(A^\dag_{2,\G k} A_{2,\G p},A^\dag_{2,\G p} A_{2,\G k};q\ri)$\footnote{In this way, the $SU_q(2)$ invariant Hamiltonian \cite{UBRIACO1996205}
\be \non
H \ = \ \sum^2_{j=1} \, \sum_{\G k} \, \varepsilon_{\G k} \, a^\dag_{j,\G k} \, a_{j,\G k} \, , 
\ee
can be rewritten as
\be \non
H \ = \  \sum_{\G k} \, \varepsilon_{\G k} \, A^\dag_{2,\G k} \, A_{2,\G k} +\sum_{\G k,\G p,\G p'} \, \varepsilon_{\G k} \, F_{\G k \G p'} \, A^\dag_{1,\G p'} \, A_{1,\G p} \, F^\dag_{\G k \G p} \, . 
\ee
In such way, incoming and outcoming particles will generally bring different momenta. In contrast, in Ref. \cite{UBRIACO1996205}, $F_{\G k \G p}= \de_{\G k \G p} (1+(q^{-1}-1)N_{2,\G k})$ and then, all particles are forced to bring the same momentum.}.
Imposing the validity of Eq.\eqref{rs}, one gets the constraint:
\be
-\sum_{\G p} \,\lf( A_{2,\G k'} \, F_{\G k \G p} \, A^\dag_{1,\G p}+ \lf[F_{\G k \G p} \, , \, A_{2,\G k'}\ri]  \, A^\dag_{1,\G p}\ri) \ = \ -q \, \sum_\G p \,  A_{2,\G k'} \,  F_{\G k \G p} \, A^\dag_{1,\G p}  \, .
\ee
This is satisfied if 
\be \label{comans}
 \lf[F_{\G k \G p} \, , \, A_{2,\G k'}\ri] \ = \ (q-1) \, A_{2,\G k'} \,  F_{\G k \G p} \, .
\ee
This can rewritten as
\be \label{ftrans}
F_{\G k \G p} \, A_{2,\G k'} \, F^{-1}_{\G k \G p} \ = \ q \, A_{2,\G k'} \, .
\ee
We thus write $F_{\G k \G p} \ = \ q^{-\mathcal{N}_{\G k \G p}}$, so that
\be
q^{-\mathcal{N}_{\G k \G p}} \, A_{2,\G k'} \, q^{\mathcal{N}_{\G k \G p}} \ = \ q \, A_{2,\G k'} \, .
\ee
For two dof $\mathcal{N}_{\G k \G p}$ must reduce to $N_2$. We now
impose the validity of Eq.\eqref{rd}:
\be
\sum_{\G p,\G p'} \, \lf( q^{-\mathcal{N}^\dag_{\G k \G p}} \, q^{-\mathcal{N}_{\G k' \G p'}} \, A_{1,\G p} \, A^\dag_{1,\G p'} + q^{-\mathcal{N}_{\G k' \G p'}} \, q^{-\mathcal{N}^\dag_{\G k \G p}} \, A^\dag_{1,\G p'} \, A_{1,\G p} \ri) \ = \ \de_{\G k \, \G k'} \, + \, (q^{-2}-1) \, A^\dag_{2,\G k'} \, A_{2,\G k} \, .
\ee
This expression can be simplified if we assume $\lf[\mathcal{N}^\dag_{\G k \G p} \, , \, \mathcal{N}_{\G k' \G p'} \ri]=0$. Then 

\be
\sum_{\G p} \,q^{-\mathcal{N}^\dag_{\G k \G p}} \, q^{-\mathcal{N}_{\G k' \G p}}  \ = \ \de_{\G k \, \G k'} \, + \, (q^{-2}-1) \, A^\dag_{2,\G k'} \, A_{2,\G k} \, .
\ee
Finally, imposing Eq.\eqref{crel} we get
\be
A_{1,\G p} \, q^{-\mathcal{N}^{\dag}_{\G k \G p}} \, A_{2,\G k'} \ = \ -\al \,  A_{2,\G k'}\, A_{1,\G p} \, q^{-\mathcal{N}^{\dag}_{\G k \G p}}+\bt \,  A_{2,\G k}\, A_{1,\G p} \, q^{-\mathcal{N}^{\dag}_{\G k' \G p}} \, .
\ee
In the case $\mathcal{N}^{\dag}_{\G k \G p}=\mathcal{N}_{\G k \G p}$, we can use Eq.\eqref{ftrans} to get
\be \label{tcon}
A_{2,\G k'} \, q^{-\mathcal{N}_{\G k \G p}} \ = \ A_{2,\G k} \, q^{-\mathcal{N}_{\G k' \G p}} 
\ee
Under the hermiticity condition of $\mathcal{N}_{\G k \G p}$, also Eq.\eqref{crel1} gives back Eq.\eqref{tcon}.

Note that, the case  $\mathcal{N}_{\G k \G p} \ = \ \mathcal{N}(A^\dag_{2,\G k} A_{2,\G p};q)$ has to be excluded. In fact, for $\G k \neq \G k'$
\be
q^{-\mathcal{N}_{\G k \G p}} \, A_{2,\G k'} \, q^{\mathcal{N}_{\G k \G p}} \ = \  A_{2,\G k'} \, .
\ee
Let us also note that the simplest ansatz $\mathcal{N}_{\G k \G p}=A^\dag_{2,\G p} A_{2,\G k}$ or $\mathcal{N}_{\G k \G p}=\ha \lf(A^\dag_{2,\G p} A_{2,\G k}+A^\dag_{2,\G k} A_{2,\G p}\ri)$ do not fit the above constraints. 
\section{Conclusions} \label{conc}
We have discussed the $SU_q(2)$ fermions, and we have shown that most of the previous applications in statistical mechanics and QFT, based on Eqs.\eqref{ut1}-\eqref{ut2}, cannot be trusted. In fact, Eqs.\eqref{ut1}-\eqref{ut2} are generally incompatible with the $SU_q(2)$ invariant relations \eqref{rd}-\eqref{rs1}. Moreover, we tried to exclude some classes of ansatz and fix plausible constraints on the form of $SU_q(2)$ fermion operators as functions of the standard ones.

Clearly, much more should be done in this direction: at the present level, we do not even know if a general solution exists or not. If a solution exists, this could represent a powerful instrument to investigate various areas of modern physics. As remarked in the introduction, $SU_q(N)$ bosons are strictly related to GUP \cite{Kempf:1993bq}. This fact represents a suggestive hint about the basic role played by quantum groups in the fundamental descriptions of nature (see also \cite{Major_1996, Noui:2002ag, Fairbairn:2010cp, Bianchi:2011uq,Jalalzadeh:2017jdo,Acquaviva:2021aau}).
\section*{Acknowledgements}
The author would like to thank A. Iorio and M. R. Ubriaco for useful discussions. The author acknowledges support from Charles University Research Center (UNCE/SCI/013).

\bibliographystyle{apsrev4-2}
\bibliography{librarySvN}

\end{document}